\documentclass[12pt,letterpaper]{article}
\usepackage[utf8]{inputenc}
\usepackage{amsmath}
\usepackage{amsfonts}
\usepackage{amssymb}
\usepackage{cite}

\usepackage{esint}
\usepackage{ulem}

\usepackage{etoolbox}
    
    \patchcmd{\maketitle}{\@fpheader}{}{}{}

\newcommand*\xbar[1]{%
  \hbox{%
    \vbox{%
      \hrule height 0.5pt 
      \kern0.3ex
      \hbox{%
        \kern-0.0em
        \ensuremath{#1}%
        \kern-0.0em
      }%
    }%
  }%
}

\newcommand{\be}{\begin{equation}}
\newcommand{\ee}{\end{equation}}
\newcommand{\bea}{\begin{eqnarray}}
\newcommand{\eea}{\end{eqnarray}}

\usepackage{braket}

\title{\boldmath  Wheeler-DeWitt equation and  Bondi-Metzner-Sachs (BMS) symmetry}

\author{Marc Henneaux\\
 {\footnotesize Universit\'e Libre de Bruxelles and 
 International Solvay Institutes,} \\ {\footnotesize Brussels, Belgium} \\ {\scriptsize and} \\
{\footnotesize Coll\`ege de France,  Universit\'e PSL, Paris, France}}

\date{}

\begin{document}

\maketitle

\begin{abstract}
The Hamiltonian formulation of the BMS symmetry on spacelike hypersurfaces enables one to define its action on solutions of the Wheeler-DeWitt equation.  Using the BRST reformulation of the theory, we provide operator expressions for the matrix elements of the BMS operators between Wheeler-DeWitt states. To that end, we construct the BRST-invariant extensions of the BMS generators, which form a BRST-extension of the BMS algebra.
\end{abstract}

\flushbottom


The Wheeler-De Witt equation \cite{DeWitt:1967yk,Wheeler:1968iap} is the central equation in the canonical approach to quantum gravity and is also an essential tool in  quantum cosmology \cite{Hartle:1983ai}. It has been the subject of renewed interest recently \cite{Raju:2019qjq,Chowdhury:2021nxw,Raju:2021lwh,Witten:2022xxp}. The connection established in  \cite{Raju:2019qjq,Chowdhury:2021nxw,Raju:2021lwh} between this equation and holography in anti-de Sitter space is very intriguing.  In order to understand how this result can be extended to flat space holography, it seems necessary to fully grasp the asymptotic properties of this equation in the flat space context and, in particular, determine how the flat space asymptotic symmetry acts on its solutions.

The Wheeler-DeWitt  equation expresses that the states $\ket{\psi}$ of the gravitational field should be annihilated by the gravitational quantum constraint operators $\hat{\mathcal H}$ and $\hat{\mathcal H}_k$, i.e., 
$\int d^3x (\zeta^\perp \hat{\mathcal H} + \zeta^k \hat{\mathcal H}_k) \ket{\psi}= 0$, 
where $\hat{\mathcal H}$ and $\hat{\mathcal H}_k$ are the quantum versions of the Hamiltonian and momentum constraint functions $\mathcal H$ and $\mathcal H_k$ \cite{Dirac:1958sc,Arnowitt:1962hi},
\be
\mathcal H =  G_{ijmn} \pi^{ij} \pi^{mn} - R \sqrt{g} \, ,  \qquad \mathcal H_k = - 2 \nabla_m{\pi_{k}}^m \, ,
\ee
($c=1$ and $16 \pi G = 1$) and where $(\zeta^\perp, \zeta^k)$ are arbitrary vector fields required to vanish at infinity.  Here, $G_{ijmn} = \frac12 g^{-\frac12}(g_{im} g_{jn} + g_{in} g_{jm} - g_{ij} g_{mn})$ is the inverse of the DeWitt supermetric.  Geometric objects are $3$-dimensional (spatial) throughout.

In the metric representation where the states $\ket{\psi}$ are functionals $\Psi[g_{ij}(x)]$ of the spatial metric $g_{ij}(x)$ ($\Psi[g_{ij}(x)] = \braket{g_{ij}(x)|\psi}$), the equation takes the familiar Wheeler-DeWitt form of a second order differential equation with respect to $g_{ij}(x)$.
Other representations are possible, where for instance the conformal geometry $\gamma_{ij} = g_{ij} g^{-\frac13}$ and the trace of the extrinsic curvature are diagonal  \cite{Dirac:1958jc} ($\Psi = \Psi[\gamma_{ij}(x), \pi(x)]$).  We shall refer to the quantum gravitational constraint equations as the ``Wheeler-DeWitt'' equation independently of the representation.

For asymptotically flat spaces, the Wheeler-DeWitt equation must be supplemented by the Schr\"odinger-like equation which gives the change of $\ket{\psi}$ when the vector field $(\xi^\perp, \xi^k)$ does not go to zero asymptotically but rather approaches an asymptotic time translation,
$i  \frac{\ket{\Psi}}{\partial T} = \hat H \ket{\psi}  $
($\hbar = 1$) 
where $\hat H$ is the Hamiltonian operator corresponding to the classical Hamiltonian
\be
H = \int d^3x(\xi^\perp \mathcal H + \xi^k \mathcal H_k) + \oint_{S_\infty} d^2 S_i (g_{ik,k} - g_{kk,i}) \, .
\ee
Here, 
$\xi^\perp \rightarrow 1 $, $ \xi^k \rightarrow 0$ for $r \rightarrow \infty$
and we have assumed asymptotically cartesian coordinates at infinity. The surface term on the sphere at infinity is the ADM energy \cite{Arnowitt:1962hi,Regge:1974zd}.

Now, asymptotic time translations form only a subgroup of the asymptotic symmetry group of gravity in the asymptotic flat context.  The complete asymptotic symmetry group was uncovered in that case by Bondi, Metzner and Sachs \cite{Bondi:1962px,Sachs:1962wk,Sachs:1962zza} -- hence the terminology ``BMS group" -- and is infinite-dimensional.  Its physical significance and implications, as well as more recent developments, are reviewed in  \cite{Strominger:2017zoo}.   
This raises the question on how the BMS symmetry acts on solutions of the Wheeler-DeWitt equation.

In order to address that question, one needs ADM-like expressions for all the BMS generators.  This problem was fully solved some time ago  \cite{Henneaux:2018hdj}, as we now briefly recall \cite{footnote-1}.

The BMS algebra is the semi-direct sum of the homogeneous Lorentz algebra parametrized by an antisymmetric tensor $\beta^{\mu \nu}$  and the supertranslations, parametrized by a function on the $2$-sphere. The spherical harmonics $l=0$ and $l=1$ correspond to the ordinary translations, while the higher spherical harmonics correspond to the pure supertranslations.  In the Hamiltonian formulation, the supertranslation parameter is naturally split into an even part $T(\theta, \varphi)$ and an odd part $W(\theta, \varphi)$.  One finds that the generator of the most general BMS transformation reads
\be
\mathring{P}_{\xi^\perp, \xi^k} = \int d^3x(\xi^\perp \mathcal H + \xi^k \mathcal H_k) + \frac12 \beta^{\mu \nu} M_{\mu \nu} + \mathcal B^{grav}_{\{T, W\}} \label{Eq:BMS0}
\ee
where the surface integrals are given in \cite{Henneaux:2018hdj,Henneaux:2019yax} (specifically, in reference  \cite{Henneaux:2019yax}, the angular momentum $M_{mn}$ is given by (6.7) or (6.36), the surface integral in the boost generators is given by (6.40) and the surface integral  $\mathcal B^{grav}_{\{T, W\}}$ is given by (6.8) or (6.37)).  The vector field $(\xi^\perp, \xi^k)$ behaves asymptotically as an infinitesimal BMS transformation (Eqs (5.5)-(5.7) of \cite{Henneaux:2019yax}),
$\xi^\perp = \beta^{ \perp i}x_i + T(\theta, \varphi) + \mathcal O\left(\frac{1}{r}\right)$, $
 \xi^k = \beta^{ki}x_i + \partial^k\left(r W(\theta, \varphi)\right) + \mathcal O\left(\frac{1}{r}\right)$.

The BMS generators are first class and real.  Provided these properties are preserved quantum-mechanically, their quantum versions unitarily map states annihilated by the constraints on states annihilated by the constraints.  The BMS group has therefore a well-defined unitary action on solutions of the Wheeler-DeWitt equations.

A difficulty with these considerations is that the naive scalar product among solutions of the constraint equations typically diverges because of the integration along the gauge directions. In order to be able to compute physical amplitudes and to talk about unitary action in the physical subspace, one needs 
 a prescription for computing scalar products among solutions of the Wheeler-DeWitt equation.
One way to tackle this problem is to fix the gauge generated by the first class constraints.  But this is not only technically complicated in the case of gravity (if it can be done at all), but also, one must make sure that the results do not depend on the gauge fixing conditions.

A better approach is given by the BRST method, which 
formally guarantees gauge independence and provides in a controlled way a scalar product that does not involve infinities related to integrations over the gauge directions.

The construction of the classical BRST charge for gravity follows the standard steps \cite{Fradkin:1975cq,Batalin:1977pb,Fradkin:1977xi,Henneaux:1985kr}.  Even though the structure functions appearing in the Poisson brackets of the constraints  are not constants (they involve the metric), the BRST charge takes the simple form valid for Lie algebras,
\begin{eqnarray}
&&\Omega^{\textrm{Min}}  = \int d^3x \big[\mathcal H(x) C^\perp(x)   +  \mathcal H_k(x) C^k(x)  \nonumber \\
&& \qquad \qquad \qquad \quad + \xbar {\mathcal P}_\perp(x) \Omega^\perp(x) +  \xbar {\mathcal P}_k(x) \Omega^k(x) \big]
\end{eqnarray}
with
\begin{equation}
 \Omega^\perp=  C^\perp_{,k} C^k \, , \qquad 
 \Omega^k = - g^{km} C^\perp C^{\perp}_{,m} - C^m C^{k}_{,m} \, 
\end{equation}
without terms of higher orders in the ghosts $(C^\perp, C^k)$ and their conjugate momenta $(\xbar {\mathcal P}_\perp, \xbar {\mathcal P}_k)$) as it would be needed for a generic theory with structure functions (instead of structure constants) in the constraint algebra: ``Gravity is a theory of rank one". We take as conventions $\{C^\mu, \xbar{\mathcal P}_\nu\}= \delta^\mu_\nu = \{\xbar {\mathcal P}_\nu, C^\mu\}$  where $\{\cdot, \cdot\}$ is the graded Poisson bracket ($\mu = \perp, k)$.  The ghosts are real and their conjugate momenta are pure imaginary, so that $(\Omega^{\textrm{Min}})^*=\Omega^{\textrm{Min}}$.  Since $\Omega^{\textrm{Min}}$ captures the proper gauge symmetries, the ghosts are required to vanish at spatial infinity.

That gravity is of rank one is not entirely obvious since the relativistic membrane of spatial dimension $n$, which has the same constraint ``algebra", is of rank $n$ \cite{Henneaux:1983um}.  In that case, the metric induced on the membrane is a composite field involving derivatives of the basic fields (the embedding coordinates), which leads to non-trivial higher-order structure functions.

The BRST charge is nilpotent of order two, $\{\Omega^{\textrm{Min}}, \Omega^{\textrm{Min}}\}=0$.  Observables are given by the BRST cohomogy at ghost number zero and reduce to standard gauge-invariant phase space functions when the ghosts are set to zero \cite{Henneaux:1988ej,Henneaux:1992ig}.  In particular, the generators of the BMS symmetry possess BRST extensions, which are most easily worked out by making the following observation.  Consider the fermionic generator $K_{\zeta^\perp, \zeta^k} = \int d^3x (\zeta^\perp \xbar {\mathcal P}_\perp +  \zeta^k \xbar {\mathcal P}_k)$ where $(\zeta^\perp, \zeta^k)$ is an arbitrary vector field that vanishes at infinity. Because $(\zeta^\perp, \zeta^k)$ goes to zero at infinity, $K_{\zeta^\perp, \zeta^k}$ is a well defined generator since the transformation it generates ($\{C^\perp, K\epsilon\} = \zeta^\perp \epsilon$, $\{C^k, K\epsilon\} = \zeta^k \epsilon$, with $\epsilon$ an infinitesimal fermionic parameter) preserves the condition that the ghosts should vanish at infinity.  Thus the (graded) Poisson bracket $A_{\zeta^\perp, \zeta^k} = \{K_{\zeta^\perp, \zeta^k}, \Omega^{\textrm{Min}}\}$ makes sense and obviously fulfills $\{A_{\zeta^\perp, \zeta^k}, \Omega^{\textrm{Min}}\}= 0$ by the Jacobi identity.  One has explicitly
\be
A_{\zeta^\perp, \zeta^k} =  \int d^3x \Big[\zeta^\perp (\mathcal H_\perp + \Delta_\perp)+  \zeta^k  (\mathcal H_k + \Delta_k) \Big] \label{Eq:BRSTTrivial}
\ee
with $\Delta_\perp = (\xbar{\mathcal P}_\perp C^k)_{,k} + \xbar{\mathcal P}_k C^\perp_{,m} g^{km} + (\xbar{\mathcal P}_k C^\perp g^{km})_{,m}$ and $\Delta_k = \xbar{\mathcal P}_\perp {C^\perp}_{,k}+ \xbar{\mathcal P}_m {C^m}_{,k} + (\xbar{\mathcal P}_k {C^m})_{,m}$.

It turns out that in verifying the BRST invariance of $A_{\zeta^\perp, \zeta^k}$, i.e., the ghost number one equation $\{A_{\zeta^\perp, \zeta^k}, \Omega^{\textrm{Min}}\}= 0$, the fact that $(\zeta^\perp, \zeta^k)$ vanishes at infinity is not needed because the vanishing of the ghosts is sufficient to ensure this property: all surface terms involved in the computation that might invalidate it are equal to zero. Thus, we can assume that $(\zeta^\perp, \zeta^k)$ goes to an asymptotic BMS transformation at infinity, and add the relevant surface integral to make $A_{\zeta^\perp, \zeta^k}$ a well-defined generator, without spoiling the BMS invariance condition.  The BRST-invariant extensions of the BMS generators (\ref{Eq:BMS0}) are therefore
\begin{eqnarray}
&& P_{\xi^\perp, \xi^k} = \int d^3x\Big[\xi^\perp (\mathcal H_\perp + \Delta_\perp)+  \xi^k  (\mathcal H_k + \Delta_k) \Big] \nonumber \\
&& \qquad \qquad \qquad \qquad \qquad + \frac12 \beta^{\mu \nu} M_{\mu \nu} + \mathcal B^{grav}_{\{T, W\}} \label{Eq:BMSExt}
\end{eqnarray}
with $(\xi^\perp, \xi^k)$ having the BMS asymptotic behaviour.  Because of the non-trivial behaviour of $(\xi^\perp, \xi^k)$ at infinity and of the presence of the surface term, these are not BRST-exact, contrary to (\ref{Eq:BRSTTrivial}). ``Proper'' gauge symmetries \cite{Benguria:1976in} have BRST-exact generators but not  improper ones.  Of course,  different continuations in the bulk of vector fields $(\xi^\perp, \xi^k)$ with the same asymptotics differ by a BRST-exact term of the form  (\ref{Eq:BRSTTrivial}), as it should be the case.  If one so wishes, one can adopt the prescription of \cite{Ashtekar:1983ug} for uniquely continuing $(\xi^\perp, \xi^k)$ in the bulk but this is not necessary.

The BRST-closed BMS generators (\ref{Eq:BMSExt}) form a BRST extension of the BMS algebra in the sense of \cite{Henneaux:1988st}, i.e., their Poisson brackets reproduce the BMS algebra up to BRST-trivial terms, which are physically irrelevant. 

Physical operators in the BRST quantum formulation of the theory are derived from the classical observables by the standard correspondence rules (we assume the quantum BRST operator to be nilpotent, $\left(\hat{\Omega}^{\textrm{Min}}\right)^2 = 0$, i.e., no gauge anomaly): they are defined by the operator BRST cohomology at ghost number zero.

The definition of the physical states is more subtle than the definition of the physical operators.  In addition to the BRST closedness condition and the identification of states that differ by a BRST exact one,
$
\hat{\Omega}^{\textrm{Min}} \ket{\psi} = 0$, $ \ket{\psi} \sim \ket{\psi} + \hat{\Omega}^{\textrm{Min}}\ket{\chi} $
which are the BRST version of the Wheeler-DeWitt equation, one must restrict the ghost number. 

To that end,  it is useful to introduce the ``non-minimal sector", which also plays a central role in the gauge fixing problem for the path integral  \cite{Fradkin:1975cq,Batalin:1977pb,Fradkin:1977xi}.  This sector, which does not change the BRST cohomology because the new variables form cohomologically trivial pairs, amounts here to putting back the Lagrange multipliers $\lambda^\perp$ and $\lambda^k$ for the constraints (the lapse and the shift) together with their conjugate momenta $b_\perp$ and $b_k$, which are constrained to vanish, $b_\perp \approx 0$, $b_k \approx 0$. The corresponding ghost pairs are denoted $(\xbar C_\perp, \mathcal P^\perp)$, $(\xbar C_k, \mathcal P^k)$ (with $\xbar C_\mu$ real and of ghost number $-1$) and the BRST operator in the extended phase takes the simple form
\be
\Omega = \Omega^{\textrm{Min}}+ i\int d^3x (b_\perp \mathcal P^\perp + b_k \mathcal P^k)  \label{Eq:BRSTComplete}
\ee
since the new constraints form an abelian algebra.  The new fields all vanish at infinity.  

With the non-minimal sector included, the BRST-Wheeler-DeWitt equation reads
\be
\hat{\Omega} \ket{\psi} = 0 \, , \qquad \ket{\psi} \sim \ket{\psi} + \hat{\Omega}\ket{\chi} \, , \label{Eq:BRSTPhysicalStates2}
\ee
with the complete BRST operator (\ref{Eq:BRSTComplete}).

One way to understand the role of the non-minimal sector follows from the fact that the ghost number operator, which is anti-hermitian, has real eigenvalues.  Hence, the scalar product of two states with definite ghost number can differ from zero only if their respective ghosts numbers add up to zero.  Therefore, if they both belong to the same subsbspace of definite ghost number, that scalar product can be different from zero only if the ghost number in question is zero. However, the states of the minimal sector with direct relationship with the Dirac quantization method have non-zero ghost number, and so, in order to extract physical amplitudes from them, one would have to deviate from the BRST general rules.  But by completing them appropriately with ghost states of the non-minimal sector, one can get zero ghost number states.  One can then strictly apply the BRST rules without having to bend them.  

Even after the BRST conditions (\ref{Eq:BRSTPhysicalStates2}) are implemented and the zero ghost condition is imposed (with non-minimal sector included), there is still in general some redundancy in the description of the physical states.  One must then impose further conditions.  There are at least three different approaches that have been developed.  We refer to \cite{Henneaux:1992ig}, chapters 14 and 16, for the details and focus right away on the approach of most interest to us, described for  general first class systems in\cite{Henneaux:1992ig}, sections 14.5 (in particular 14.5.3 through 14.5.5) and 16.5.  We formulate it here in the gravity case.

In that approach, on imposes, in addition to (\ref{Eq:BRSTPhysicalStates2}) and the ghost number zero condition, that the states should be annihilated by the ghosts 
\be
\hat C^\mu \ket{\psi} = 0 \, , \qquad \hat{\xbar C}_\mu \ket{\psi} = 0 \, , \qquad \mu = (\perp, k) \, . \label{Eq:UpStates}
\ee
We thus write
$
\ket{\psi} = \Psi[g_{ij}, \lambda^\mu]\ket{\uparrow}  
$
where  $\ket{\uparrow}$ is the maximally filled state annihilated by the ghosts (denoted by $\psi_{C=0, \bar C= 0}$ in \cite{Henneaux:1992ig}). The BRST invariance condition imposes $\frac{\delta \Psi}{\delta \lambda^\mu(x)} =0$, i.e., the physical states should not depend on the Lagrange multipliers, but their dependence on the metric is completely unrestricted. Thus we have
\be
\ket{\psi} = \Psi[g_{ij}]\ket {0}\ket{\uparrow}  \label{Eq:StatesUseful}
\ee
where  $\ket{0}$  stands  for the state $\ket{b_\mu =0}$ (denoted by $\psi_{b=0}$ in \cite{Henneaux:1992ig}) with zero eigenvalue  for the momenta conjugate to the Lagrange multipliers, which is indeed such that $\braket{\lambda^\mu \vert 0}$ is independent of $\lambda^\mu$.  

Although $\Psi[g_{ij}]$ is subject to no equation, one recovers the Wheeler-DeWitt equations through the quotient relation $\ket{\psi} \sim \ket{\psi} + \hat{\Omega}\ket{\chi}$ \cite{footnote0}.  Thus, one can fully capture the physics without having to solve the Wheeler-DeWitt equation. The use of these convenient states is particularly  advocated (for the $\perp$-sector) in \cite{Chandrasekaran:2022cip,Witten:2022xxp}.

The subspace of (\ref{Eq:BRSTPhysicalStates2}) defined by (\ref{Eq:UpStates}) is invariant under the action of the BMS generators (\ref{Eq:BMSExt}).  This is because the commutators of the ghost fields with these generators are proportional to the ghosts \cite{footnote2}.  The results from the BMS representation theory are then applicable to that invariant subspace (see \cite{Bekaert:2025kjb} for recent developments on BMS representations as well as references to earlier work).

The states  (\ref{Eq:StatesUseful}) enable one to construct matrix elements of gauge-invariant observables between physical states.  More precisely, the so-called projected kernel of the gauge-invariant observable $A_0$  is given, in the representation where the metric is diagonal, by the matrix elements (we drop the hat over operators when no confusion is possible)
\be
A_0^P [g_{ij}^{(2)}, g_{ij}^{(1)}] =  \braket{g_{ij}^{(2)},0 ,\uparrow \vert A \exp{[K,\Omega]} \vert  g_{ij}^{(1)},0 ,\uparrow} \, ,  \label{Eq:ProjKernel0}
\ee
(graded commutator, i.e., in this case, anticommutator) where $A$ is a BRST-invariant extension of $A_0$ and the ``gauge-fixing" fermion $K$ is appropriately chosen, a task performed in the next paragraphs.  Here, the states $\ket{g_{ij}^{(1)}}$ are eigenstates of the metric operator with eigenvalues $g_{ij}^{(1)}(x)$, i.e., in the metric reprentation, these states are given by $\Psi[g_{ij}] =\prod_x \delta(g_{ij}(x) - g_{ij}^{(1)}(x))$ (just like $\ket{q^i_0}$ is given by $\prod_i \delta(q^i-q^i_0)$ in the representation where $\hat{q}^i$ is diagonal). The projected kernel $A_0^P [g_{ij}^{(2)}, g_{ij}^{(1)}]$ is defined in  \cite{Henneaux:1992ig} and is the kernel of $A_0$ in a basis of physical states expressed as functions of the metric \cite{footnote3}. It contains the information about all the scalar products of $A_0$ between physical states.  The scalar product of two functionals $f$, $g$ of the metric, the Lagrange multipliers and the ghosts $C$, $\xbar C$ is the naive one, $\braket{f| g} =  \int \mathcal D g_{ij} \mathcal D\lambda \mathcal D C  \mathcal D \xbar C f^*[g_{ij}, \lambda, C, \xbar C] g[g_{ij}, \lambda, C, \xbar C]$.

Without the insertion of the operator $\exp{[K,\Omega]}$, which is BRST equivalent to the identity, the scalar product 
$\braket{g_{ij}^{(2)},0 ,\uparrow \vert  g_{ij}^{(1)},0 ,\uparrow}$ (for $A = I$) is ill-defined because it involves an integral over the Lagrange multipliers (on which the states do not depend and which is thus infinite) times $\braket{\uparrow \vert  \uparrow}$, which is zero.  

The gauge-fixing fermion $K$ must be chosen so as to regularize that scalar product 
by eliminating $0 \times \delta(0)$. The resulting value formally does not depend on the choice of $K$ due to the properties of the BRST formalism. 
A convenient choice of $K$ is 
$
K = \int d^3x (\lambda^\perp \xbar {\mathcal P}_\perp + \lambda^k \xbar {\mathcal P}_k)
$
so that
\begin{equation}
 [K, \Omega] =  \int d^3x (i \lambda^\mu \mathcal{H}_\mu - \xbar {\mathcal P}_\mu \mathcal P^\mu + i \lambda ^\mu \Delta_\mu)
\end{equation}
where $(\mu = \perp, k)$.
The matrix elements (\ref{Eq:ProjKernel0}) read then
\begin{eqnarray}
&& \hspace {-.4cm} A_0^P [g_{ij}^{(2)}, g_{ij}^{(1)}] = \int  \mathcal D\lambda^\mu \braket{g_{ij}^{(2)}\uparrow \vert A e^{ \int d^3 x \rho(\lambda)}|g_{ij}^{(1)}\uparrow}  \label{Eq:GroupAverage} \\
&&
\rho (\lambda) = i \sigma(\lambda)- \bar {\mathcal P}_\mu \mathcal P^\mu \, , \quad \sigma(\lambda) =  \lambda^\mu \mathcal{H}_\mu  +  \lambda^\mu \Delta_\mu 
\end{eqnarray}
where we have exhibited the functional integral over the Lagrange multipliers.  The zero factor due to $\braket{\uparrow  \vert \uparrow} = 0$ disappears because of the presence of the operator $ \bar {\mathcal P}_\mu \mathcal P^\mu$ in the exponential.

Because ${C^k}_{\perp \perp}$ involves the metric, one cannot decouple the ghost contribution from the metric contribution. 
It can be verified, however, that for constraints $G_\mu \approx 0$ forming a Lie algebra, the integrand in the scalar product of the states ($A=I$) factorises as $\braket{f| e^{i \lambda^\mu G_\mu} |g} \mu(\lambda)$ where $\mu(\lambda)$ is the invariant (Haar) measure  on the corresponding group. 
 Hence, in that case, the (regularized) scalar product reduces to the group average,
$
 \int  d\lambda^\mu \mu(\lambda) \braket{f| e^{i \lambda^\mu G_\mu} |g} \, 
$
which was considered in \cite{Teitelboim:1983fk,Teitelboim:1983fj,Higuchi:1991tm}.    This also shows that the integral over the multipliers is not infinite any more, because the states on which $e^{i \lambda^\mu G_\mu}$ acts are not annihilated by the constraints and have, in fact, finite integral along the gauge orbits, as indicated in \cite{footnote0}.

The BRST method of \cite{Henneaux:1992ig}, based on a definition of the physical states involving the ghost states $\ket{\uparrow}$,  generalizes therefore the ``group average technique'', or ``refined algebraic quantization method'' for handling constraints forming a Lie algebra developed  in \cite{Ashtekar:1995zh,Giulini:1998rk,Marolf:2008hg} and which plays an important role in the recent works \cite{Chandrasekaran:2022cip,Kaplan:2024xyk} (for related BRST considerations, see \cite{Barvinsky:1993jf,Batalin:1994rd,Shvedov:2001ai}).

A particular case of (\ref{Eq:GroupAverage}) is obtained by taking $A = P_{\xi^\perp, \xi^k}$, in which case one gets the projected kernel of the BMS generator $\mathring{P}_{\xi^\perp, \xi^k}$
$$
 \mathring{P}_{\xi^\perp, \xi^k}^P [g_{ij}^{(2)}, g_{ij}^{(1)}] = \int  \mathcal D\lambda^\mu  \braket{g_{ij}^{(2)}\uparrow \vert P_{\xi^\perp, \xi^k} \, e^{ \int d^3 x \rho(\lambda)}|g_{ij}^{(1)}\uparrow}
$$
It is instructive to consider the projected kernel of the BMS group element $e^{i P_{\xi^\perp, \xi^k}}$, which involves the product of the exponentials
$$
 e^{i \big[ \int d^3 x \sigma(\xi) +  \frac12 \beta^{\mu \nu} M_{\mu \nu} + \mathcal B^{grav}_{\{T, W\}}\big]} \, e^{ \int d^3 x \rho(\lambda)}
$$
Using the Baker-Campbell-Hausdorff formula, one can rewrite the product of exponentials as
$$
e^{ \int d^3 x \rho(N) + \frac{i}{2} \beta^{\mu \nu} M_{\mu \nu} + i \mathcal B^{grav}_{\{T, W\}}+ \cdots}
$$
where the dots  stands for the contributions from the higher order commutators and where $N^\mu = \lambda^\mu + \xi^\mu$.  The variables $N^\mu$ have the same asymptotic behaviour as $\xi^\mu$.  If $\xi^\mu$ where to vanish at infinity, one knows that modulo trivial terms, that exponential would just be 
$
e^{ \int d^3 x \rho(N)} 
$
as it would follow by simply taking $K = \int d^3x N^\mu \xbar{\mathcal P}_\mu$.  Thus, the contributions from the multiple commutators would yield trivial terms. But these multiple commutators involve at least one $\lambda^\mu$ or a ghost variable, which all vanish at infinity, and so, are insensitive to the asymptotic behaviour of $\xi^\mu$.  It follows that these multiple commutators also contribute trivial terms when $\xi^\mu$ does not vanish at infinity, leading to the following expression for the projected kernel of the BMS group element $e^{iP_{\xi^\perp, \xi^k}}$,
\begin{equation}
 \hspace{-.4cm} \int  \mathcal DN^\mu \braket{g_{ij}^{(2)}\uparrow \vert e^{ \int d^3 x \rho(N) +  \frac{i}{2} \beta^{\mu \nu} M_{\mu \nu} +i \mathcal B^{grav}_{\{T, W\}}} |g_{ij}^{(1)}\uparrow} ,
\end{equation}
by making the change of variables $N^\mu = \lambda^\mu + \xi^\mu$.  One must integrate aver all $N^\mu(x)$ that have the asymptotic behaviour dictated by the given group element.  Because of the non-trivial asymptotic behaviour of $N^\mu$ and of the surface terms, the matrix elements of the exponential of the improper (BMS) gauge symmetries differ from the identity,  contrary to what would be the case for the exponentials of proper gauge symmetries.

In the case where the BMS group element belongs to the Poincar\'e subgroup, this expression reproduces the one given in \cite{Teitelboim:1983fi} (with the product of $\exp{[\int d^3x \lambda^\perp \xbar{\mathcal P}_\perp,\Omega]} $ times $ \exp{[\int d^3x \lambda^k \xbar{\mathcal P}_k,\Omega]}$ replacing $\exp{[\int d^3x \lambda^\mu \xbar{\mathcal P}_\mu,\Omega]}$ to make the match).
But because we are interested in expectation values of observables between solutions of the Wheeler-DeWitt equation, and not in defining a causal propagator suitable for perturbation theory, the restriction $N^\perp>0$ of \cite{Teitelboim:1981ua} is not imposed here.

While we have not attempted to go beyond the formal operator level in this paper, which puts instead the emphasis on conceptual aspects, a satisfactory definition of the various BRST expressions could presumably be reached by perturbative developments as in \cite{Chowdhury:2021nxw,Witten:2022xxp}.

With a full control of the BMS symmetry in the Wheeler-De Witt context, it would be interesting to investigate how the connection between holography and solutions of the Wheeler-DeWitt equation established in 
 \cite{Chowdhury:2021nxw,Raju:2019qjq} for anti-de Sitter gravity extends to the asymptotically flat case (see also \cite{Marolf:2008mf,Jacobson:2019gnm}).  It is hoped to return to this question in the future.
In this regard, it might be necessary to include along exactly the same lines the logarithmic supertranslations \cite{Fuentealba:2022xsz,Fuentealba:2023syb}, which provide a canonical description of the BMS Goldstone fields \cite{Strominger:2017zoo} at spatial infinity.

\section*{Acknowledgments}
The author is grateful to Edward Witten for insightful comments.  This work was partially supported by FNRS-Belgium (convention IISN 4.4503.15), as well as by research funds from the Solvay Family.






\begin{thebibliography}{99}
\bibitem{footnote2}
For this argument to work, it is necessary to have  $C^k \ket{\psi} = 0$ if  $C^\perp \ket{\psi}=0$.  Note also that as defined by (\ref{Eq:BMSExt}), the BMS generators do not act on the variables of the non-minimal sector.  One might consider a non-zero action by modifying (\ref{Eq:BMSExt}) in a way that preserves the BRST-invariance condition, e.g., by adding the trivial term $\{\int d^3x (\mathcal P^\perp \pi_\perp+ \mathcal P^i \pi_i), \Omega\}$.  This does not change the key property that the subspace defined by  (\ref{Eq:UpStates}) is BMS invariant.
\end{thebibliography}
\end{document}